# Bremsstrahlung radiation power in non-Maxwellian plasmas

Chaotong Yang[1], Kai Li[1,*] and Huasheng Xie[2,3,*]

[1] Centre for Theoretical and Computational Physics, College of Physics, Qingdao University, Qingdao, 266071, People's Republic of China
[2] Hebei Key Laboratory of Compact Fusion, Langfang 065001, People's Republic of China
[3] ENN Science and Technology Development Co., Ltd, Langfang 065001, People's Republic of China

E-mail: kaili@qdu.edu.cn and xiehuasheng@enn.cn

**Abstract**

In plasmas, bremsstrahlung includes electron-ion (e-i) bremsstrahlung and electron-electron (e-e) bremsstrahlung. Bremsstrahlung radiation power loss is one of the most significant losses in fusion plasmas, which is more pronounced in higher temperature fusion. The factors that affect bremsstrahlung power include the mean electron energy and the electron velocity distribution shape. In this study, we systematically study the influence of the electron velocity distribution shape on the bremsstrahlung power with fixed total electron energy. It was found that the existing electron velocity distribution shapes have little effect on the bremsstrahlung power. In addition, by analyzing the bounds of bremsstrahlung power, we have provided the theoretical upper and lower bounds of e-i radiation. Our analysis reveals that the e-i bremsstrahlung power depends critically on the degree of energy distribution concentration. Specifically, in non-relativistic regimes, concentrated energy distributions enhance the radiation power, whereas in high-temperature relativistic regimes, such concentration suppresses it. This discrepancy arises from the distinct contributions of high-energy electron populations to radiation power across different energy regimes. For e-e bremsstrahlung, a similar dependence on energy concentration is observed. Furthermore, e-e radiation power exhibits additional sensitivity to the anisotropy of the electron velocity distribution function. These rules could provide a basis for reducing bremsstrahlung power losses in fusion plasmas.

## 1. Introduction

Bremsstrahlung radiation, generated by the deceleration of charged particles within Coulomb fields, represents a ubiquitous form of electromagnetic emission in high-temperature plasma environments [1, 2], particularly within nuclear fusion systems. As a dominant energy dissipation mechanism in high-temperature plasmas, bremsstrahlung radiation poses a fundamental challenge to the realization of advanced fusion fuels such as proton-boron-11 (p-B11). For p-B11 fusion, where electron temperatures are relativistic (~100-300 keV), bremsstrahlung losses dominate the energy balance and threaten the feasibility of net energy gain [3-5]. Recent studies have proposed suppressing these losses by engineering non-thermal electron velocity distributions, such as introducing energy cutoffs to redistribute high-energy electrons [6], which motivates this work.

In non-Maxwellian fusion plasmas, the ion velocity distribution function primarily governs fusion reactivity, while the electron velocity distribution function dominates energy dissipation through bremsstrahlung radiation. To enhance net energy output, improvements in fusion reactivity must be synergistically combined with precise control of electron energy losses. Recent studies have investigated the impact of characteristic non-Maxwellian ion distribution shapes (e.g., Bi-Maxwellian and slowing-down distributions) on fusion reactivity and established upper bounds for reactivity enhancement [7-9]. Non-Maxwellian electron distributions have been demonstrated to exist in fusion plasmas [10-14]. While certain aspects of bremsstrahlung radiation in non-Maxwellian distributions have been studied [15-18], systematic investigations into how electron velocity distribution shapes influence bremsstrahlung losses remain absent.

A pivotal assumption in the analysis of nonequilibrium fusion plasmas, notably in Rider's seminal work [3,19], is that the functional dependence of bremsstrahlung radiation power on the mean electron energy is approximately independent of the electron velocity distribution shape. This assumption underpins simplified models of radiative losses in non-Maxwellian plasmas but remains unverified.

Recent advancements in bremsstrahlung modeling further underscore the necessity of reevaluating this assumption. Munirov and Fisch [6] demonstrated that relativistic bremsstrahlung cross-sections exhibit a pronounced dependence on high-energy electron populations, suggesting that distribution shape (particularly the suppression of superthermal electrons) may significantly alter total radiation. Meanwhile, Xie [20] developed high-accuracy analytical formulas for bremsstrahlung power, thereby enabling precise quantification of subtle distribution-dependent effects previously obscured by modeling inaccuracies.

This paper addresses this unresolved question by systematically evaluating the impact of electron velocity distribution shapes (including energy cutoff, beam-like, and anisotropic distributions) on bremsstrahlung power in fusion plasmas. Using validated framework from Ref. [20], we quantify deviations from Rider's assumption and assess their implications for p-B11 fusion scenarios. Our results reveal that while certain specific shapes of the velocity distribution function can influence bremsstrahlung power, the actual deviations are relatively slight. Therefore, bremsstrahlung power remains primarily determined by the mean electron energy. However, our further theoretical analysis demonstrates that Rider's assumption holds only for weakly non-thermal distributions. Specifically, in some extreme cases, the radiation power significantly depends on parameters beyond the mean electron energy.

The paper is organized as follows. In Section 2, we calculate the bremsstrahlung radiation power for non-Maxwellian electron distributions and compared these results with Maxwellian distributions possessing identical mean electron energy. In Section 3, the dependence of bremsstrahlung power on the electron velocity distribution shape was analyzed through a simplified theoretical framework. Finally, we summarize our findings in Section 4.

## 2. Bremsstrahlung radiation power under various distributions

In this section, we will calculate the radiation power for electron-ion (e-i) bremsstrahlung and electron-electron (e-e) bremsstrahlung under various distributions. Our approach is to first calculate the radiation power of the Maxwellian distribution, and for non-Maxwellian distributions, we mainly focus on the ratio of its radiation power to that of a Maxwellian distribution with the same mean electron energy. Below, we introduce the computational models and formulas used in this work.

The bremsstrahlung radiation power per unit volume is given by:

$$P = \frac{n_1 n_2}{1+\delta_{12}} \iiint h\nu \boldsymbol{p_{12}} \frac{d\sigma}{d\nu} f_1(\boldsymbol{p_1}) f_2(\boldsymbol{p_2}) \, d\boldsymbol{p_1} d\boldsymbol{p_2} d\nu \quad (1)$$

where $\boldsymbol{p_{12}}$ is the relative velocity of two particles, $n_1, n_2$ is the number density of two kinds of particles, $f_1(\boldsymbol{p_1}), f_2(\boldsymbol{p_2})$ is the distribution of two kinds of particles, $\sigma$ is the cross-section, $h$ is Planck's constant, $\nu$ is the frequency of the generated photon, $\boldsymbol{p_1}, \boldsymbol{p_2}$ is the momentum of two kinds of particles.

It can be seen that the cross section plays a crucial role in the calculation of bremsstrahlung. In the calculation of e-i radiation power, we respectively adopt the cross sections in the Sommerfeld model [21] and the Bethe-Heitler model [22] in the non-relativistic and relativistic rigimes.

To study the radiation power under different distribution functions, it is convenient to integrate over the photon energy and use fitting formulas to simplify the calculations. We define:

$$W_E(E_0) = \int_0^{E_0} \frac{\nu\sqrt{3}\pi}{AR_y Z^2} \frac{d\sigma}{d\nu} dh\nu \quad (2)$$

where $Z$ is the ion charge number.
Combining equations (1), (2), and the properties of e-i radiation, we obtain:

$$P = n_e n_i \frac{AR_y Z^2 c}{\sqrt{3}\pi} \int_0^\infty (2e)^{1/2} W_E(e) f(e) de \quad (3)$$

for the non-relativistic regime, and:

$$P = n_e n_i \frac{AR_y Z^2 c}{\sqrt{3}\pi} \int_0^\infty \frac{p}{e+1} W_E(e) f(e) de \quad (4)$$

for the relativistic regime. Here, $e = \frac{E}{m_e c^2}$ is the reduced electron kinetic energy (not the electron charge).

Ref. [20] provides the following expression for $W_E(e)$:

$$W_E(e) = c_0 f_{nr} f_r (1-f_z) \quad (5)$$

with

$$f_{nr} = 1 + c_{nr1}\left\{1 - exp\left[-\left(c_{nr2}\frac{Z^2}{e}\right)^{c_{nr4}}\right]\right\}$$
$$- \left(c_{nr1} + 1 - \frac{1}{c_0}\right) exp\left[-\left(\frac{e}{c_{nr3}Z^2}\right)^{c_{nr5}}\right]$$

$$f_r = 1 + 1.069\left(\frac{p}{\varepsilon}\right)^2 - 2.1939\left(\frac{p}{\varepsilon}\right)^3$$
$$+ \left[2.315\left(\frac{p}{\varepsilon}\right)^2 - 1.5652\left(\frac{p}{\varepsilon}\right)^4\right]\varepsilon[ln(p+\varepsilon) - \frac{1}{3}]$$

$$f_z = \frac{0.433Z\left(\frac{10e}{\sqrt{Z}}\right)^{-1.106}}{exp\left[2.8975\left(\frac{e}{\sqrt{Z}}\right)^{-0.256}\right] - 1}$$

Here $\varepsilon = E/m_e c^2$, $p = P/m_e c$, $e = E_k/m_e c^2$ are respectively the reduced electron total energy, the reduced electron momentum, and the reduced electron kinetic energy.

Using equations (3), (4) and (5), we can quickly calculate the radiation power for arbitrary distribution.

For e-e radiation power, Ref. [23] provides a formula for direct calculation in the center-of-mass system:

$$P_{ee} = \frac{n_e^2 m_e c^3 r_e^2 \alpha}{2} \iint_{-\infty}^{+\infty} \frac{\varepsilon_1 + \varepsilon_2}{\varepsilon_1 \varepsilon_2} \sqrt{\frac{1}{2}[(\boldsymbol{p_1 p_2}) - 1]}$$
$$Q_{cm}(\varepsilon_{cm}) f(\boldsymbol{p_1}) f(\boldsymbol{p_2}) d\boldsymbol{p_1} d\boldsymbol{p_2} \quad (6)$$

where $\boldsymbol{p_1 p_2} = \varepsilon_1 \varepsilon_2 - \boldsymbol{p_1} \cdot \boldsymbol{p_2}$, $\varepsilon_{cm} = \sqrt{(\boldsymbol{p_1 p_2}+1)/2}$

For Maxwellian distributions, Refs. [24] and [25] provide analytical formulas:

$$P_{ee} = \frac{4n_e^2 m_e c^3 r_e^2 \alpha}{t\left[K_2\left(\frac{1}{t}\right)\right]^2} \int_1^\infty \varepsilon_{cm}^2 p_{cm}^2 K_2\left(\frac{2\varepsilon_{cm}}{t}\right) Q_{cm}(\varepsilon_{cm}) d\varepsilon_{cm} \quad (7)$$

where $p_{cm} = \sqrt{(\boldsymbol{p_1 p_2}-1)/2}$, $K_2$ is the modified Bessel function of the second kind, $t = k_B T/(m_e c^2)$ is the reduced temperature.

In our calculations, we use the new fitting formula for $Q_{cm}$ provided in Ref. [20]:

$$Q_{cm}(\varepsilon) = 8\frac{p^2}{\varepsilon}F_{ee}\left\{1.007 + 2.11\frac{p}{\varepsilon} - 3.45\left(\frac{p}{\varepsilon}\right)^2 + 2\left[1 + 2.12\left(\frac{p}{\varepsilon} - 1\right)\right]ln(\varepsilon + p)\right\} \quad (8)$$

with

$$F_{ee}(\varepsilon) = \frac{1}{2}(tanh\{0.62[\log_{10}(\varepsilon - 1) + 4.96]\} + 1)$$

Refs. [6] and [23] also provide an approximate two-dimensional integral:

$$P_{ee} = n_e^2 m_e c^3 r_e^2 \alpha \iint \frac{\varepsilon_1 + \varepsilon_2}{4\varepsilon_1 \varepsilon_2 p_1 p_2} J(\varepsilon_1, \varepsilon_2) f(\varepsilon_1) f(\varepsilon_2) d\varepsilon_1 d\varepsilon_2 \quad (9)$$

with

$$J(\varepsilon_1, \varepsilon_2) = \left(\frac{u}{2} - 2\right)\sqrt{u^2 - 1} - \frac{11}{12}u^2 + \frac{20}{3}u - \frac{8}{3}ln(u+1)$$
$$+ \left[\frac{3}{2} + \left(\frac{u}{2} - \frac{8}{3}\frac{u+2}{u+1}\right)\sqrt{u^2 - 1}\right]ln(u + \sqrt{u^2 + 1}) + \frac{7}{4}ln^2(u + \sqrt{u^2 - 1})\Big|_{\varepsilon_1\varepsilon_2 - p_1 p_2}^{\varepsilon_1\varepsilon_2 + p_1 p_2}$$

This formula has a maximum error of about 5% compared to the analytical formula for Maxwellian distributions.

In calculating e-e radiation, we are more concerned with high-temperature radiation, as e-e radiation power is relatively small at low temperatures. Additionally, due to the complexity of some non-Maxwellian distributions, Monte Carlo methods will be used to calculate the radiation power for certain distribution functions.

## 2.1 Bremsstrahlung radiation power for Maxwellian distribution and monoenergetic distribution

The Maxwell-Boltzmann (M-B) distribution in the non-relativistic regime is given by:

$$f(e) = \frac{2}{\sqrt{\pi}} t^{-\frac{3}{2}} \sqrt{e} \exp\left(-\frac{e}{t}\right) \quad (10)$$

The Maxwell-Jüttner (M-J) distribution in the relativistic regime is given by:

$$f(e) = \frac{(e+1)\sqrt{e^2 + 2e}}{tK_2\left(\frac{1}{t}\right)} \exp\left(-\frac{e+1}{t}\right) \quad (11)$$

Using equations (3), (5), and (10), we can calculate the non-relativistic e-i radiation power. Similarly, using equations (4), (5), and (11), we can calculate the relativistic e-i radiation power. Note that for non-relativistic calculations, we only use the non-relativistic terms in the expressions. For e-e radiation power, we use equations (6), (7), and (8). The results of these calculations are shown in figure 1. The figure provides a clear visual representation of the relative intensities of distinct radiation mechanisms across varying temperature regimes.

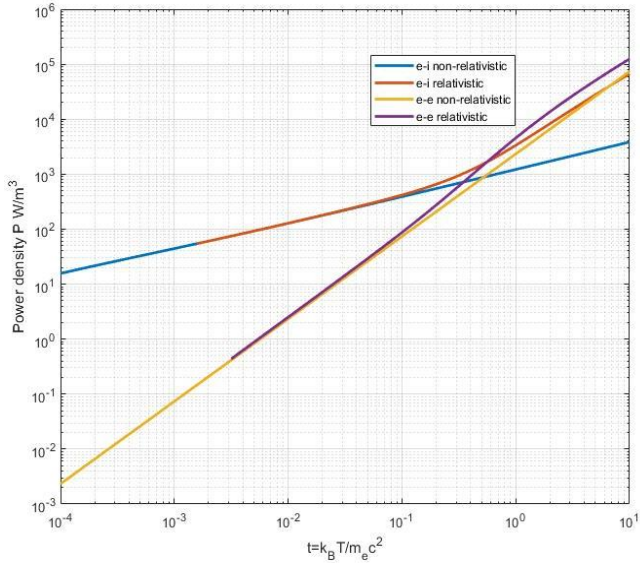

**Figure 1.** The Maxwellian plasmas bremsstrahlung power, where we take $Z = 1$ and $n_e = 10^{19} \text{m}^{-3}$.

Consider an isotropic monoenergetic distribution:

$$f(e) = \delta(e - e_c) \quad (12)$$

We can calculate the radiation power caused by this distribution. The result is shown in figure 2 and figure 3.

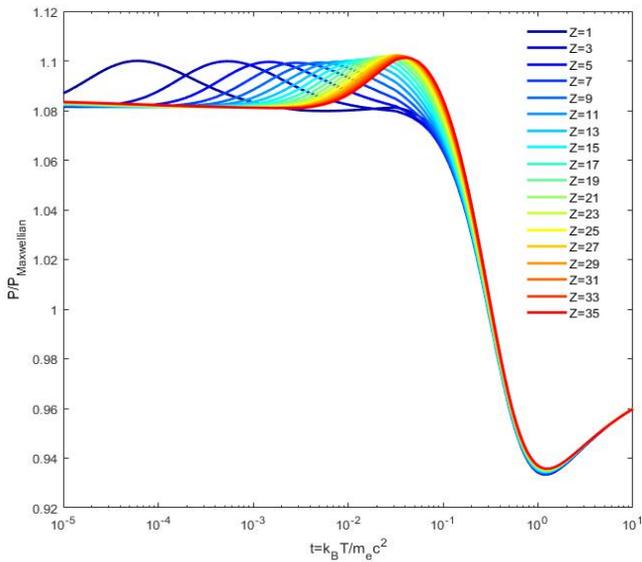

**Figure 2.** The e-i bremsstrahlung power ratio of the delta function to Maxwellian distribution at different temperatures and Z values.

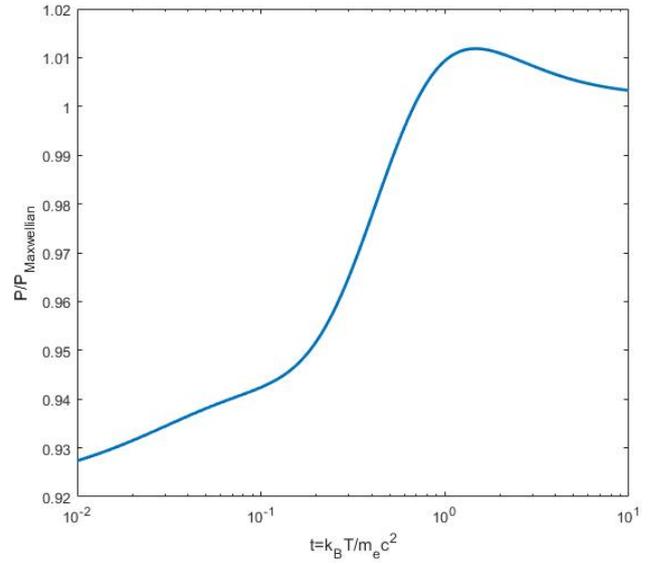

**Figure 3.** The e-e bremsstrahlung power ratio of the delta function to Maxwellian distribution at different temperatures.

We can see that the power deviation between the monoenergetic distribution and the Maxwellian distribution is around 10%.

In the subsequent calculations and analyses, the e-i bremsstrahlung is modeled with the atomic number $Z = 1$ by default.

## 2.2 Bremsstrahlung radiation power for energy cutoff distribution

In Ref. [6], the energy cutoff effect of relativistic plasma is calculated. The distribution is:

$$f(e) = \begin{cases} N_{CONST} f_{Maxwellian}(e), & e < e_{max} \\ 0, & e > e_{max} \end{cases} \quad (13)$$

We also calculated this effect. Figure 4 and figure 5 compares the computational results reproduced from Ref. [6] with those obtained using methodologies in this work.

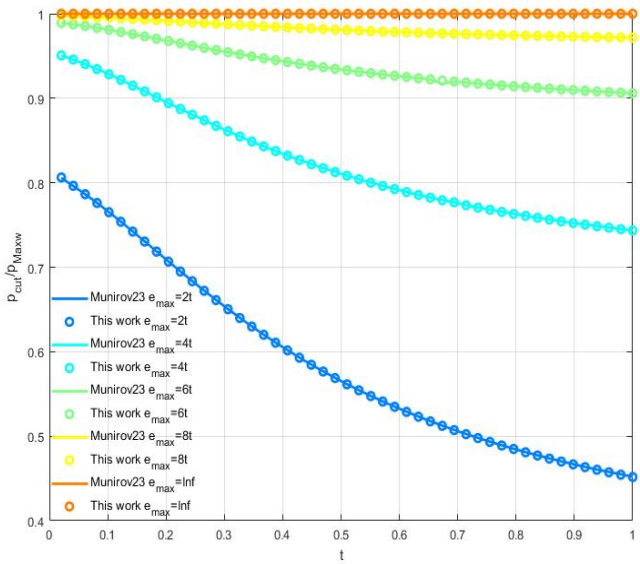

**Figure 4.** The e-i bremsstrahlung power ratio of the different energy cut-off distributions to Maxwellian distribution at different temperatures. The solid lines denote results reproduced from Ref. [6], while the dots represent computational outcomes obtained in this study.

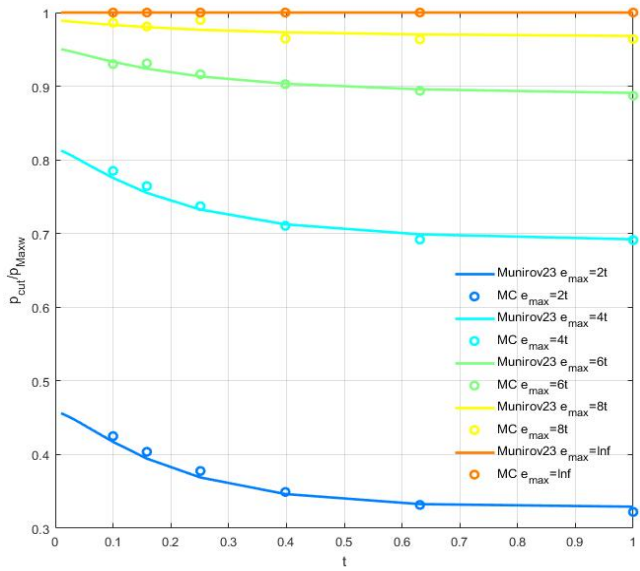

**Figure 5.** The e-e bremsstrahlung power ratio of the different energy cut-off distributions to Maxwellian distribution at different temperatures. The solid lines denote results reproduced from Ref. [6], while the dots represent computational outcomes obtained via the Monte Carlo method in this study.

The result what is actually obtained is the difference in bremsstrahlung power between the distribution function obtained after the redistribution of the superthermal electrons into lower energies and the original Maxwellian distribution function. Therefore, the cause of the final result actually includes the influence of the variation of the mean electron energy and the variation of the shape of the electron distribution function. We recalculate this effect while keeping the mean energy of the two distribution functions equal with equation (3), (4), (5) and (13), and give the results at low temperatures for e-i radiation power. The result is shown in figure 6 and figure 7. We can see that the power difference caused by the change of electron velocity distribution shape is still not significant.

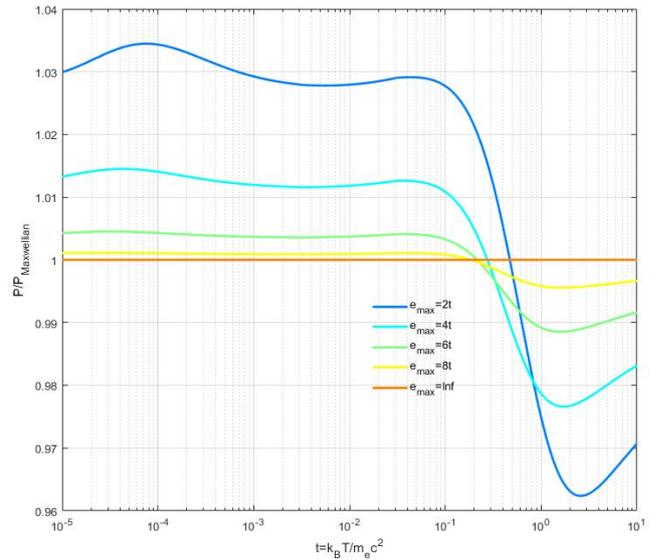

**Figure 6.** The e-i bremsstrahlung power ratio of the different energy cutoff distributions to Maxwellian distribution at different temperatures when Z=1.

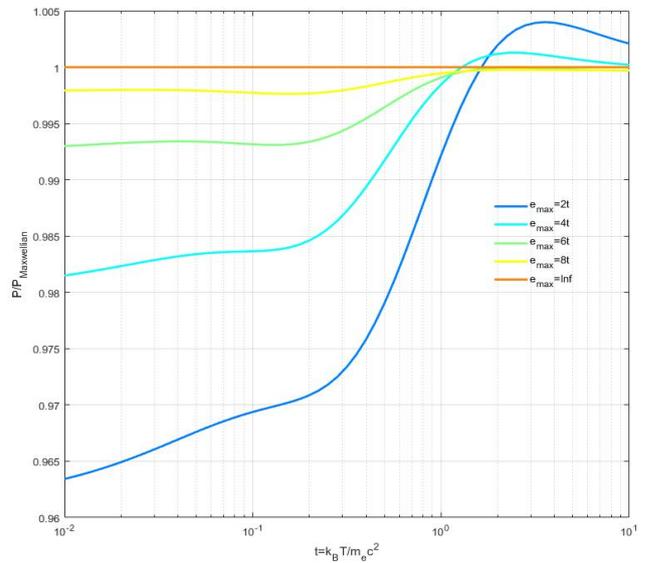

**Figure 7.** The e-e bremsstrahlung power ratio of the different energy cutoff distributions to Maxwellian distribution at different temperatures.

*2.3 Bremsstrahlung radiation power for super-Gaussian distribution*

Langdon [26] showed that nonlinear effects in inverse bremsstrahlung absorption result in a distortion of the electron distribution function, leading to super-Gaussian distributions. And these distributions are measured directly in experiments performed on the UCD AURORA II device [27].

In accordance with Ref. [28, 29], the fitted distribution is:

$$f_m(x,v,t) = C_m exp[-(v/v_m)^m] \quad (14)$$

with

$$v_m^2 = \frac{3kT}{M_e}\frac{\Gamma(5/m)}{\Gamma(3/m)} \text{ and } C_m = \frac{N_e}{4\pi}\frac{m}{\Gamma(3/m)v_m^3}$$

This can be written in the form we need:

$$f(e,t) = \frac{m}{2\Gamma(3/m)}\left(\frac{2\Gamma(5/m)}{3\Gamma(3/m)}\right)^{\frac{3}{2}}\sqrt{e}\,t^{-\frac{3}{2}}$$

$$\exp\left(-\left(\frac{e}{t}\frac{2\Gamma(5/m)}{3\Gamma(3/m)}\right)^{\frac{m}{2}}\right) \quad (15)$$

where

$$m = 2\sim 5$$

When $m=2$, the distribution is Maxwell-Boltzmann distribution. In these distributions, it is worth noting that distribution functions with different values of $m$ have nearly the same mean electron kinetic energy as long as they are at the same $t$ value. We can then calculate the e-i bremsstrahlung power ratio of the distribution of different $m$ values to the Maxwellian distribution with equation (3) and equation (15). The result is shown in figure 8. And we can calculate the e-e bremsstrahlung power ratio by Monte-Carlo method. The result is shown in figure 9.

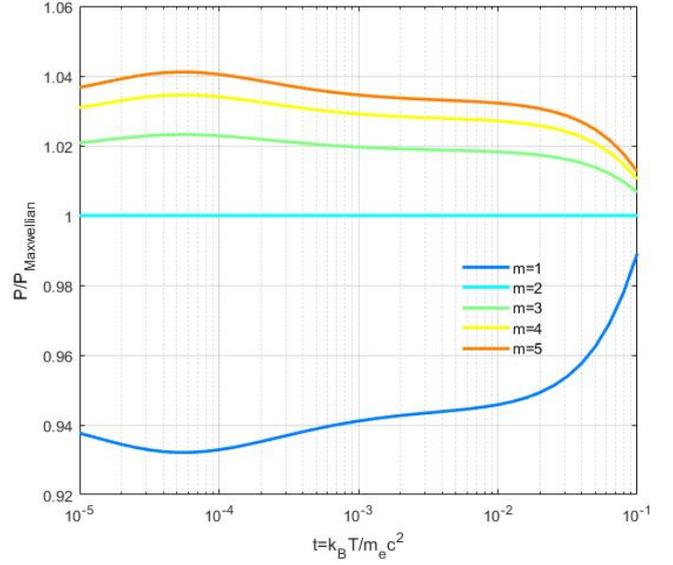

**Figure 8.** The e-i bremsstrahlung power ratio of the different super-Gaussian distributions to Maxwellian distribution at different temperatures.

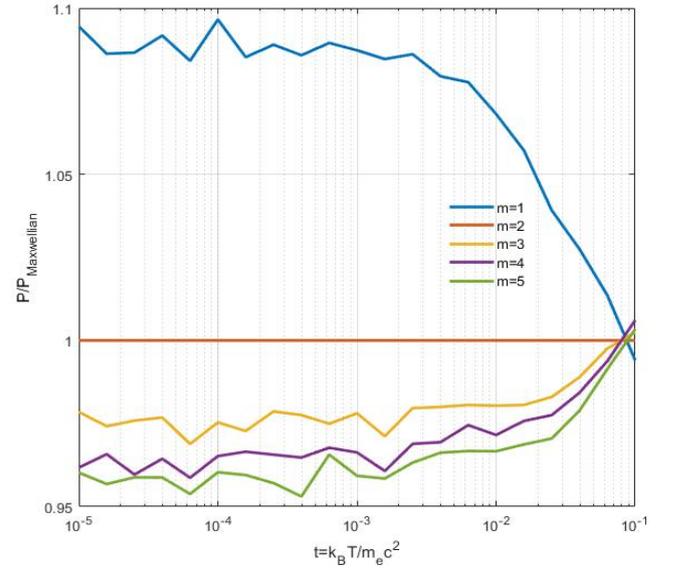

**Figure 9.** The e-e bremsstrahlung power ratio of the different super-Gaussian distributions to Maxwellian distribution at different temperatures, sampling points = 500,000.

We can see that under the influence of these distribution function shapes, the energy loss caused by e-i bremsstrahlung increases by several percent, and this increase is proportional to the $m$ value, which is opposite to e-e bremsstrahlung power. But overall, the difference caused by the shape of these distribution functions is still very limited. It should be noted that the distribution of $m=1$ does not really exist, we just calculate it numerically.

## 3. Theoretical analysis of distribution function effects

In the previous section, we calculated the radiation power for different specific distribution functions. In this section, we aim to analyze the distribution function effects with fixed total energy. And note that all calculations of e-e bremsstrahlung in this section only consider relativistic electrons.

*3.1 The bounds of e-i bremsstrahlung radiation power*

For e-i bremsstrahlung, we can first express the electron energy distribution function as a superposition of delta functions:

$$f(e) = \frac{1}{N}\sum_{n=1}^{N} \delta(e - e_n) \quad (16)$$

and the constraint is:

$$\frac{1}{N}\sum_{n=1}^{N} e_n = e_{Maxw} \quad (17)$$

As $N \to \infty$, this form can be used to approximate arbitrary distribution functions.

We first investigate the case with a finite value of $N$, where the bremsstrahlung power is numerically evaluated by randomly selecting energy values that satisfy the fixed total energy constraint. The non-relativistic and relativistic results for varying $N$ are shown in figure 10 and figure 11.

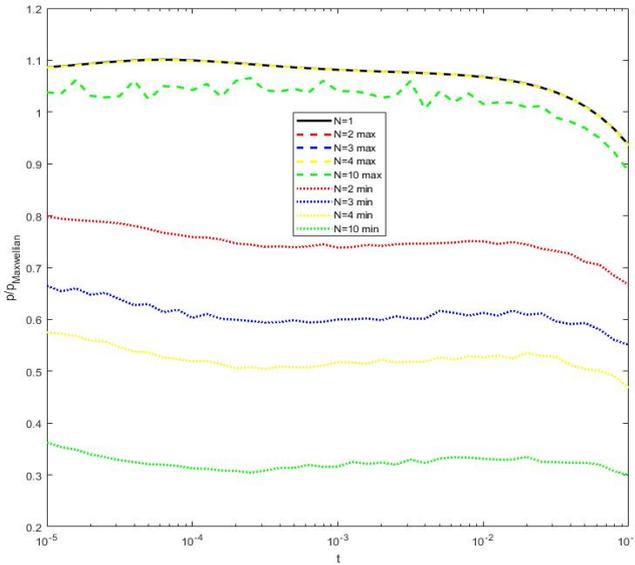

**Figure 10.** The upper and lower bounds of non-relativistic e-i bremsstrahlung power derived from stochastic energy sampling, sampling points = 10,000.

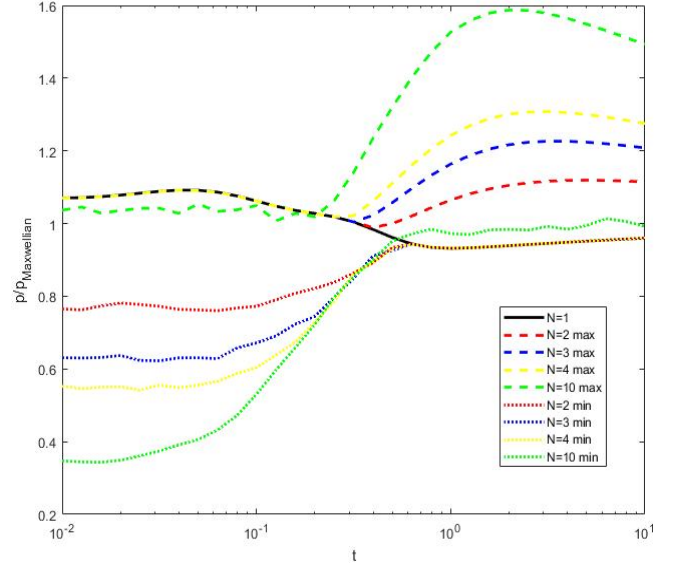

**Figure 11.** The upper and lower bounds of relativistic e-i bremsstrahlung power derived from stochastic energy sampling, sampling points = 10,000.

It can be observed that for non-relativistic plasmas, $N = 1$ invariably serves as the upper bound of radiative power. As the $N$ increases, the lower bound gradually decreases. In the relativistic plasmas, the behavior aligns with that of non-relativistic plasmas at lower temperatures. However, under higher-temperature conditions, $N = 1$ instead becomes the lower bound of power, while the upper bound progressively elevates with increasing $N$.

Let $N_i$ denote the occurrence count of energy $e_i$ in Equation (16), define $x_i = \frac{N_i}{N}$. Then, Equation (16) can be reformulated as:

$$f(e) = \sum_{n=1}^{N} x_n \delta(e - e_n) \quad (18)$$

the constraint is:

$$\sum_{n=1}^{N} x_n \delta(e - e_n) = e_{Maxw} \quad (19)$$

$$\sum_{n=1}^{N} x_n = 1 \quad (20)$$

and the objective function is:

$$p = \sum_{n=1}^{N} x_n p_n \quad (21)$$

Where $p_n$ denotes the radiation power induced by a mono-energetic particle beam, which has been calculated in section 2.

We can utilize the Lagrange multiplier method to determine the extrema of equation (21) while adhering to the constraint given in equation (19) and (20). We define the Lagrangian as follows:

$$L = \sum_{n=1}^{N} x_n p_n - \lambda \left[ \sum_{n=1}^{N} x_n \delta(e - e_n) - e_{Maxw} \right] \quad (22)$$

For the aforementioned expression to attain extrema, the following necessary conditions must be satisfied:

$$g(e, e_{Maxw}) = \frac{\partial p_n}{\partial e_n} = \lambda \quad (23)$$

We have calculated the values of the non-relativistic $g$ and revealed its monotonic decreasing nature. This implies that, under such conditions, only the case of $n = 1$ (i.e., a mono-energetic distribution) can satisfy Equation (23). For relativistic plasmas, while the low-energy regime aligns with non-relativistic electron behavior, two distinct energy values emerge in the high-energy regime that fulfill the constraints, indicating the existence of second extremum points. For boundary points (specifically corresponding to particle beam energies approaching zero), we primarily compute the boundary conditions for $n = 2$.

The rationale for selecting $n = 2$ is justified as follows. First, by analyzing the derivative of the objective function, it can be demonstrated that regardless of the value of $N$, there exists only one extremum point corresponding to the case where all energy components are equal. Boundary points, by definition, occur when one energy component approaches zero. Consequently, the boundary for $n = 3$ effectively reduces to a line segment analogous to the $n = 2$ case. Crucially, not all points along this line segment represent minima, the configuration attaining the minimum value aligns with the $N = 2$ scenario. Thus, the $N = 3$ problem is reducible to the $N = 2$ framework. By extension, this methodology generalizes to arbitrarily large $N$, as any high-dimensional boundary condition can be similarly mapped to the $N = 2$ paradigm. For instance, a minimum at $x_1 = 0.0001$ inherently corresponds to the $N = 10000$ case.

When adopting distinct values of the parameter $x_1$, the functional values at these boundary points exhibit parametric dependence. By systematically varying $x_1$, we generate a series of boundary evaluations. Subsequently, the global maximum and minimum radiation power under these constraints can be determined by comparing these points. Additionally, we observe that when $x_1$ takes small values, the boundary value approaches the maximum value very closely. Therefore, we use the boundary value as a substitute for the value of the second extremum point. The relationship between $e_1$ and radiation power at different temperatures is shown in figure 12. The upper and lower bounds of e-i bremsstrahlung power is shown in figure 13 and figure 14.

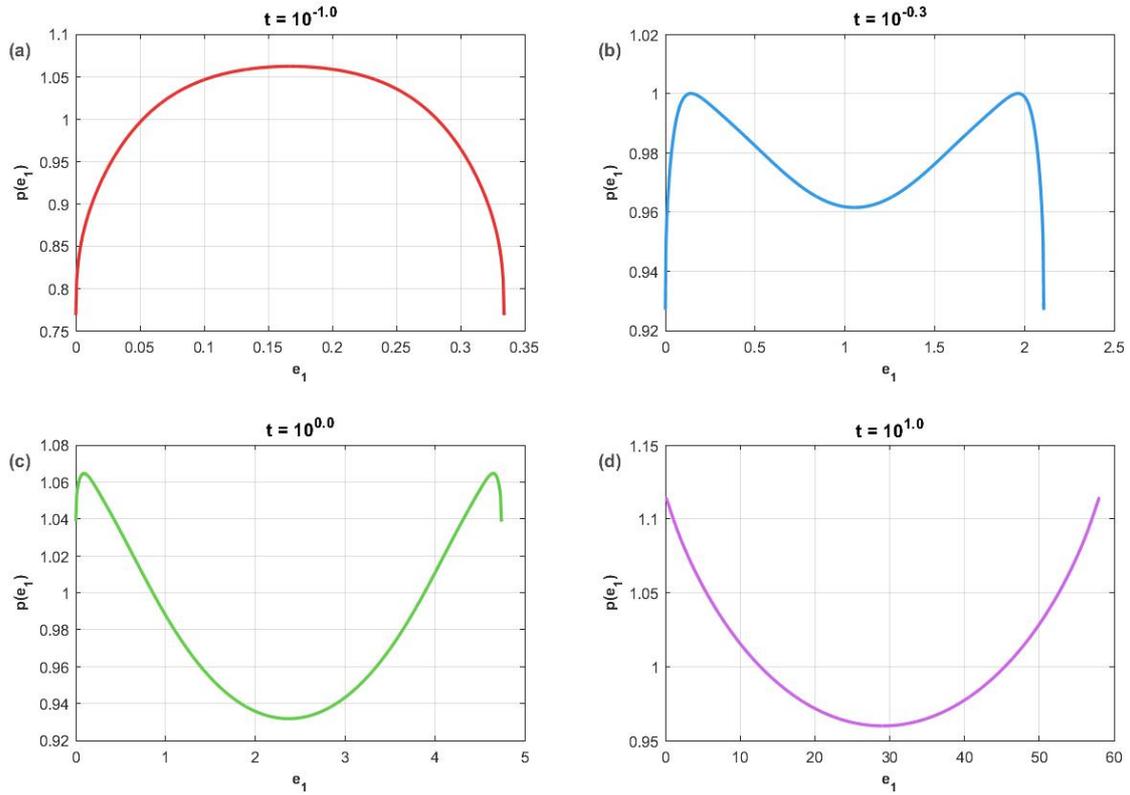

**Figure 12.** The relationship between $e_1$ and radiation power at different temperatures when $x_1 = 0.5$: (a) non-relativistic and low-temperature relativistic cases, (b), (c) transition temperature case, (d) high-temperature relativistic case.

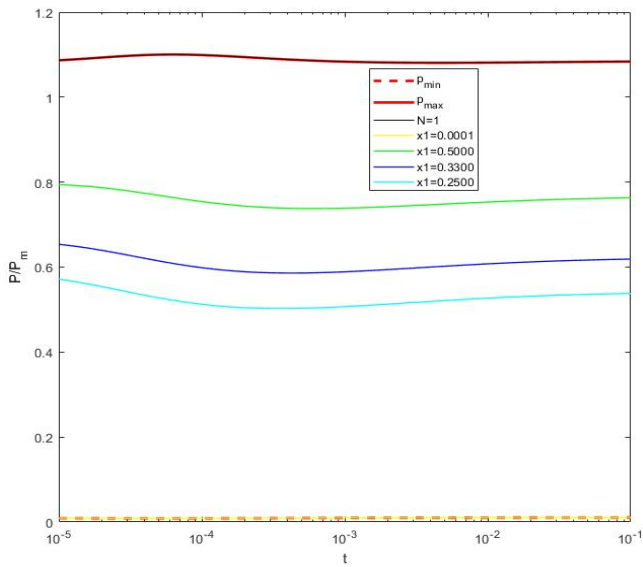

**Figure 13.** The upper and lower bounds of non-relativistic e-i bremsstrahlung power when $x_1 \in (0.0001, 0.9999)$.

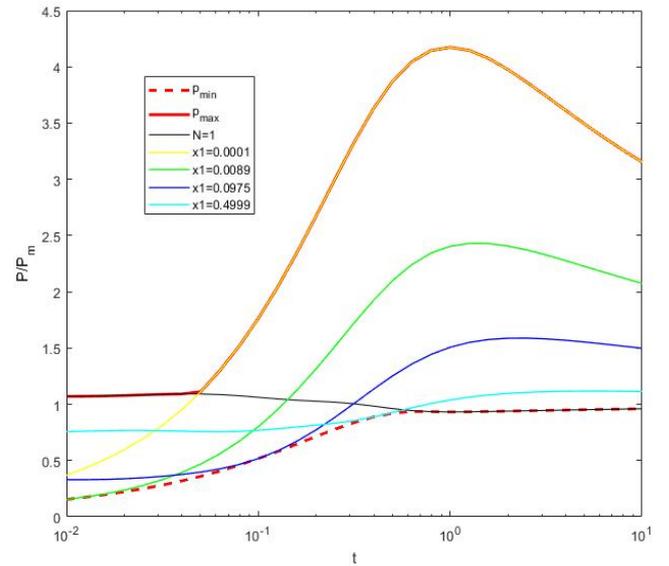

**Figure 14.** The upper and lower bounds of relativistic e-i bremsstrahlung power when $x_1 \in (0.0001, 0.9999)$.

*3.2 Calculations of e-e bremsstrahlung radiation power*

For e-e bremsstrahlung radiation, deriving results as rigorous as those for e-i bremsstrahlung remains challenging. Nevertheless, meaningful insights can still be obtained through analogous analytical frameworks. We can first express the electron momentum distribution as a superposition of delta functions:

$$f(\vec{p}) = \frac{1}{N}\sum_{n=1}^{N}\delta(\vec{p}-\vec{p}_n) \quad (24)$$

and the constraint is:

$$\frac{1}{N}\sum_{n=1}^{N}\sqrt{\vec{p}_n^{\,2}+1} = \varepsilon_{Maxw} \quad (25)$$

Now, We randomly choose $\vec{p}_n$ to find the maximum and minimum of equation (6), while satisfying equation (25). The result is shown in figure 15. For the case of $N = 2$, the functional values under varying inter-particle angles ($\theta$) and energy ratios ($\varepsilon_1/\varepsilon_2$) are shown in figure 16. Additionally, We assign specific weights to each beam ($x_1$) and calculate the radiation power accordingly. The result is shown in figure 17.

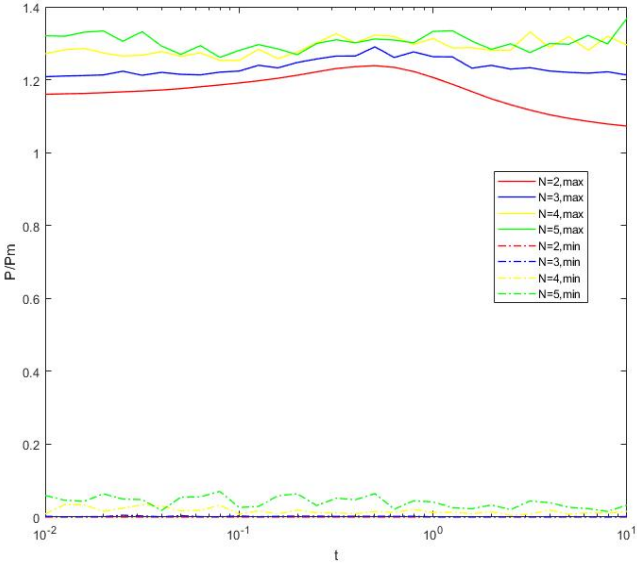

**Figure 15.** The upper and lower bounds of e-e bremsstrahlung power derived from stochastic energy sampling, sampling points = 100,000.

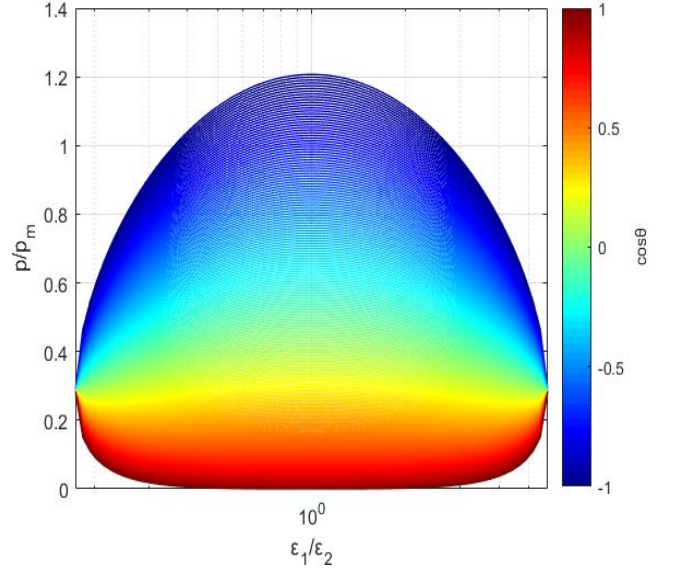

**Figure 16.** Relativistic e-e bremsstrahlung power under different energy distributions with $\cos\theta \in (-1,1)$ and $t = 1$, where the change from blue to red corresponds to $\cos\theta \in (-1,1)$.

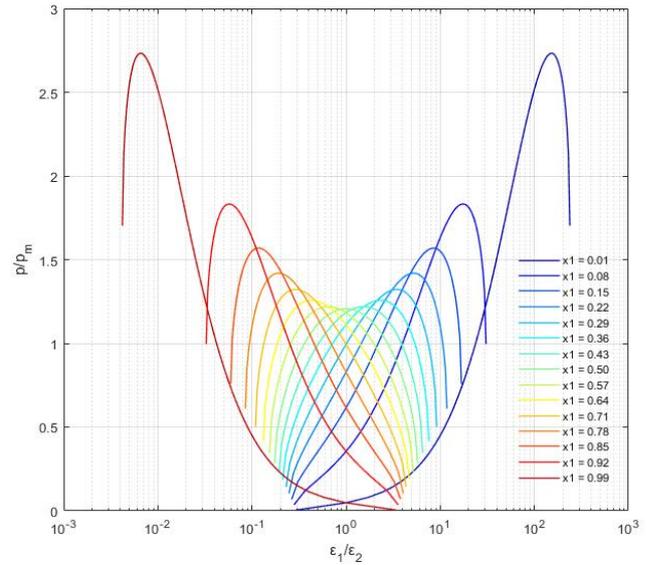

**Figure 17.** Relativistic e-e bremsstrahlung power under different energy distributions with $x_1 \in (0.01,0.99)$, $t = 1$ and $\cos\theta = -1$.

*3.3 Analysis and discussion*

As evident from figure 13, the radiation power upper limit for $N = 1$ corresponds to non-relativistic plasmas, while the lower limit asymptotically approaches zero, consistent with the trend observed in figure 10 from random sampling. Notably, the energy distribution at vanishing radiation power exhibits a characteristic where most electrons possess negligible kinetic energy, while a minority carry significant energy. This indicates that energy distribution imbalance reduces bremsstrahlung power in non-relativistic electron.

Figure 14 demonstrates that for relativistic electrons, the behavior aligns with non-relativistic cases at low temperatures ($t < 0.1$), but inversely diverges at high temperatures ($t > 1$). This implies that a small population of high-energy electrons dominates radiation in relativistic regimes, as further visualized in figure 12. Although derived from discrete energy distributions, these conclusions remain generalizable to continuous distribution functions. This provides critical insight: energy distribution inhomogeneity induces radiation power variations at fixed mean electron energy. Specifically:

Non-relativistic regime: Concentrated energy distributions enhance radiation power.

Relativistic regime ($t > 1$): Energy concentration suppresses radiation power.

Transition zone : Intermediate behavior emerges.

Two test distributions are:

$$f(e,t) = af_{Maxwellian}(e,t) + (1-a)f_{Maxwellian}(e,bt) \quad (26)$$

$$f(e,t) = \frac{1}{\sigma\sqrt{2\pi}} exp\left(-\frac{(e - e_{Maxw}(t))^2}{2\sigma^2}\right) \quad (27)$$

By varying parameters $b$ and $\sigma$, we quantify radiation power dependence on energy variance at fixed mean energy. Computational results under non-relativistic and relativistic frameworks (figures 18-19) confirm our predictions.

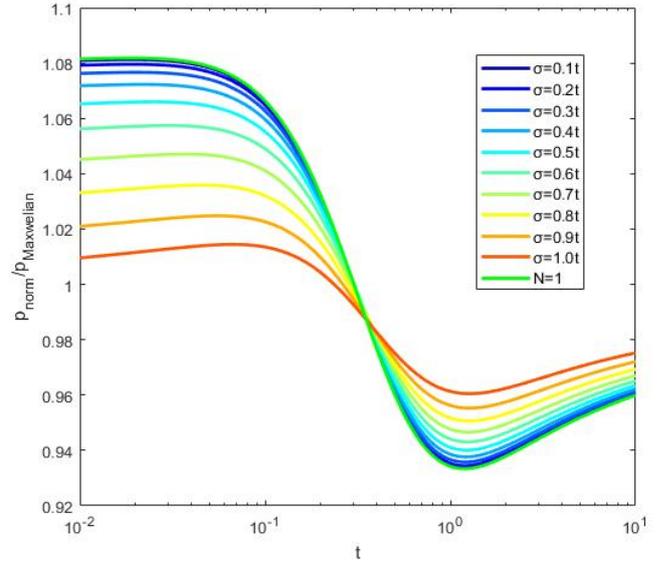

**Figure 19.** Relativistic e-i bremsstrahlung power under equation (27).

Our computational results demonstrate a qualitative correlation between energy concentration and radiation power. We propose employing energy variance to characterize this relationship. As shown in figure 10, this approach may hold for both non-relativistic electrons and relativistic electrons at high temperatures.

We further verify that distribution functions with identical mean electron energy and energy variance yield nearly identical radiation power (see figure 20).

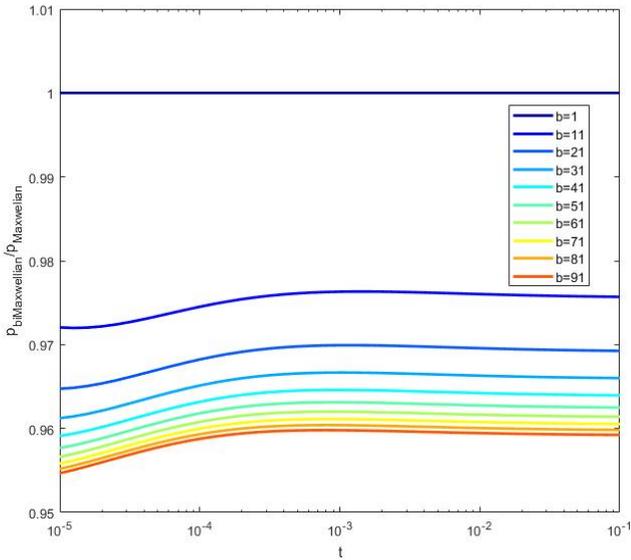

**Figure 18.** Non-relativistic e-i bremsstrahlung power under equation (26) with $a = 0.1$.

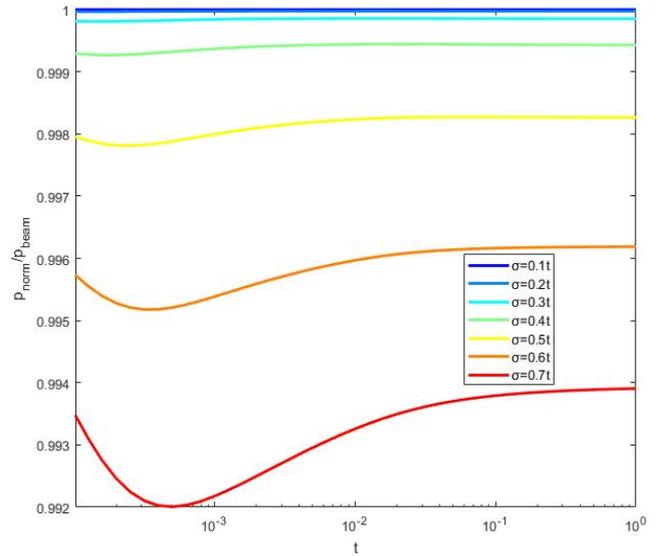

**Figure 20.** Non-relativistic regimes, the power ratio between Gaussian distributions and two-beam distributions possessing identical mean energy and energy variance.

When solely considering energy distributions, relativistic e-e and e-i radiation power demonstrate certain similarities. Specifically, high-energy electrons contribute significantly to both radiation mechanisms. Nevertheless, our primary interest lies in investigating the influence of velocity anisotropy in distribution functions on e-e radiation power. As illustrated in figure 16, e-e radiation exhibits strong dependence on the relative velocities between electrons.

First, consider an ensemble of Maxwellian-distributed electrons with momentum vectors isotropically oriented. We then modify the system by preserving individual electron speed magnitudes while enforcing alignment of all momentum vectors within a specific angular range. This manipulation preserves the energy distribution while altering the momentum-space anisotropy. The resultant difference in e-e radiation power quantifies the impact of momentum distribution ordering on radiative processes. The computational results are presented in figure 21.

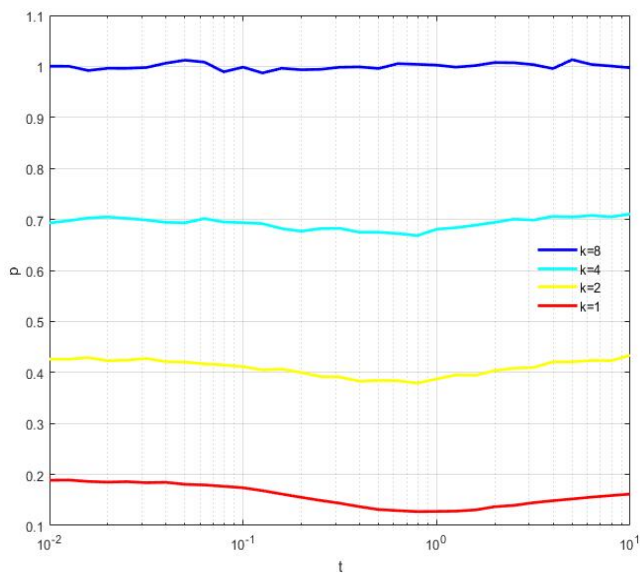

**Figure 21.** Relativistic e-e bremsstrahlung power, where the parameter $k = 8, 4, 2, 1$ represents the distribution of momentum across 8, 4, 2 and 1 quadrants, sampling points = 40,000.

## 4. Summary and conclusion

This work systematically investigates the influence of distribution functions on bremsstrahlung power in fusion plasmas. Through the study of specific distribution functions, it is found that for different distribution functions with the same mean electron energy, the bremsstrahlung power is relatively close, with deviations from the Maxwellian distribution being less than 10%.

Through theoretical calculations and analysis, we have determined the upper and lower bounds of e-i radiation power in both relativistic and non-relativistic regimes. Furthermore, we point out that in the non-relativistic and high-temperature relativistic cases, the variance of the velocity distribution function can be used as a metric to quantify the degree of its influence on radiation power. In isotropic distributions, e-e radiation exhibits characteristics similar to e-i radiation. Additionally, a notable feature of e-e radiation is that anisotropic distribution functions strongly affect the radiation power. As shown in this paper, it is possible to significantly reduce radiation power by making the electron momentum distribution more ordered. However, sustaining such non-equilibrium distributions over extended durations necessitates continuous external energy input, casting uncertainty on their potential to achieve net energy gain. Additionally, the pronounced anisotropy inherent to these distribution functions may induce plasma instabilities, further complicating their practical implementation.

Our study robustly validates and refines Rider's hypothesis. For practical fusion plasma applications, the mean electron energy emerges as the dominant factor governing bremsstrahlung power, with deviations induced by neglecting specific electron velocity distribution shapes typically remaining below 10%. Furthermore, our theoretical analysis rigorously establishes that this hypothesis holds under the condition that the variance of the energy distribution is not excessively large. For e-e bremsstrahlung, an additional requirement of plasma isotropy must be satisfied.

It should be noted that the conclusions of this study are predicated upon existing theoretical models, and their validity hinges on the precision of these models within specific application scenarios. For instance, as extrapolated from the conclusions presented in this work, an electron velocity distribution function exhibiting a hot tail effect would reduce e-i bremsstrahlung radiation for non-relativistic electrons while enhancing e-i bremsstrahlung radiation from relativistic electrons. In practical applications, particular attention must be paid to the energy regime of the high-energy electron tail, with conclusions selected accordingly based on this critical parameter.

A more rigorous analysis of e-e radiation power may be the future work. Furthermore, in the longer term, the characteristics of bremsstrahlung power from additional types of interactions—such as electron-positron bremsstrahlung—may emerge as subjects of future research.

## Acknowledgements

This work was supported by the National Natural Science Foundation of China under Grant No. 12205157.